\documentclass[letterpaper]{article}
\usepackage{aaai}
\usepackage{times}
\usepackage{helvet}
\usepackage{courier}
\usepackage[utf8]{inputenc}
\usepackage[pdftex]{graphicx}  
\usepackage{array}
\usepackage[hyphens]{url}
\usepackage{multirow}
\usepackage{tabularx}
\usepackage{subfigure}
\usepackage{amsmath}
\usepackage{longtable}
\usepackage{dblfloatfix}
\usepackage{balance}
\usepackage[table]{xcolor}
\usepackage{balance}

\begin{document}

	\title{When Politicians Talk:\\Assessing Online Conversational Practices of Political Parties on Twitter}

	 \setlength\titlebox{2.5in} 
	 \author{
	 Haiko Lietz\\
	 GESIS \& University of Duisburg-Essen\\
	 Cologne and Duisburg, Germany\\
	 haiko.lietz@gesis.org\\
	 \And
	 Claudia Wagner\\
	 GESIS \& University of Koblenz\\
	 Cologne and Koblenz, Germany\\
	 claudia.wagner@gesis.org\\
	 \AND
	 Arnim Bleier\\
	 GESIS - Leibniz Institute for the Social Sciences\\
	 Cologne, Germany\\
	 arnim.bleier@gesis.org\\
	 \And
	 Markus Strohmaier\\
	 GESIS \& University of Koblenz\\
	 Cologne and Koblenz, Germany\\
	 markus.strohmaier@gesis.org\\
	 }

	\maketitle

	\begin{abstract}
		Assessing political conversations in social media requires a deeper understanding of the underlying practices and styles that drive these conversations. In this paper, we present a computational approach for assessing online conversational practices of political parties. Following a deductive approach, we devise a number of quantitative measures from a discussion of theoretical constructs in sociological theory. The resulting measures make different -- mostly qualitative -- aspects of online conversational practices amenable to computation. We evaluate our computational approach by applying it in a case study. In particular, we study online conversational practices of German politicians on Twitter during the German federal election 2013. We find that political parties share some interesting patterns of behavior, but also exhibit some unique and interesting idiosyncrasies. Our work sheds light on (i) how complex cultural phenomena such as online conversational practices are amenable to quantification and (ii) the way social media such as Twitter are utilized by political parties.

	\end{abstract}

	\begin{figure*}[ht!]
		\centering
		\vspace{-15pt}
		\begin{tabular}{ccc}
			\subfigure[Following ($H$=$0.83$)]{\includegraphics[width=0.30\textwidth]{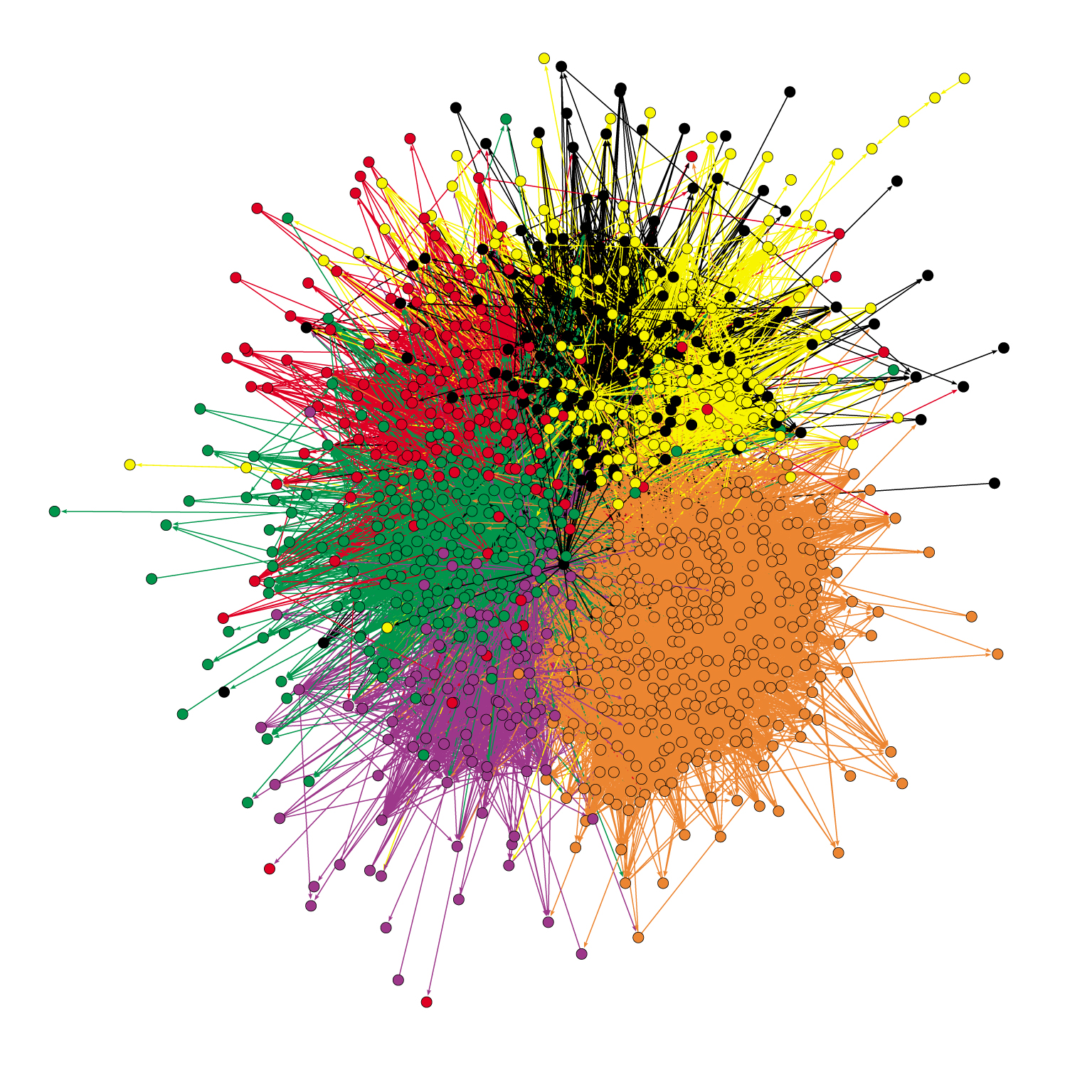}
			\label{hashtagNetwork}}
			\subfigure[Retweeting ($H$=$0.90$)]{\includegraphics[width=0.30\textwidth]{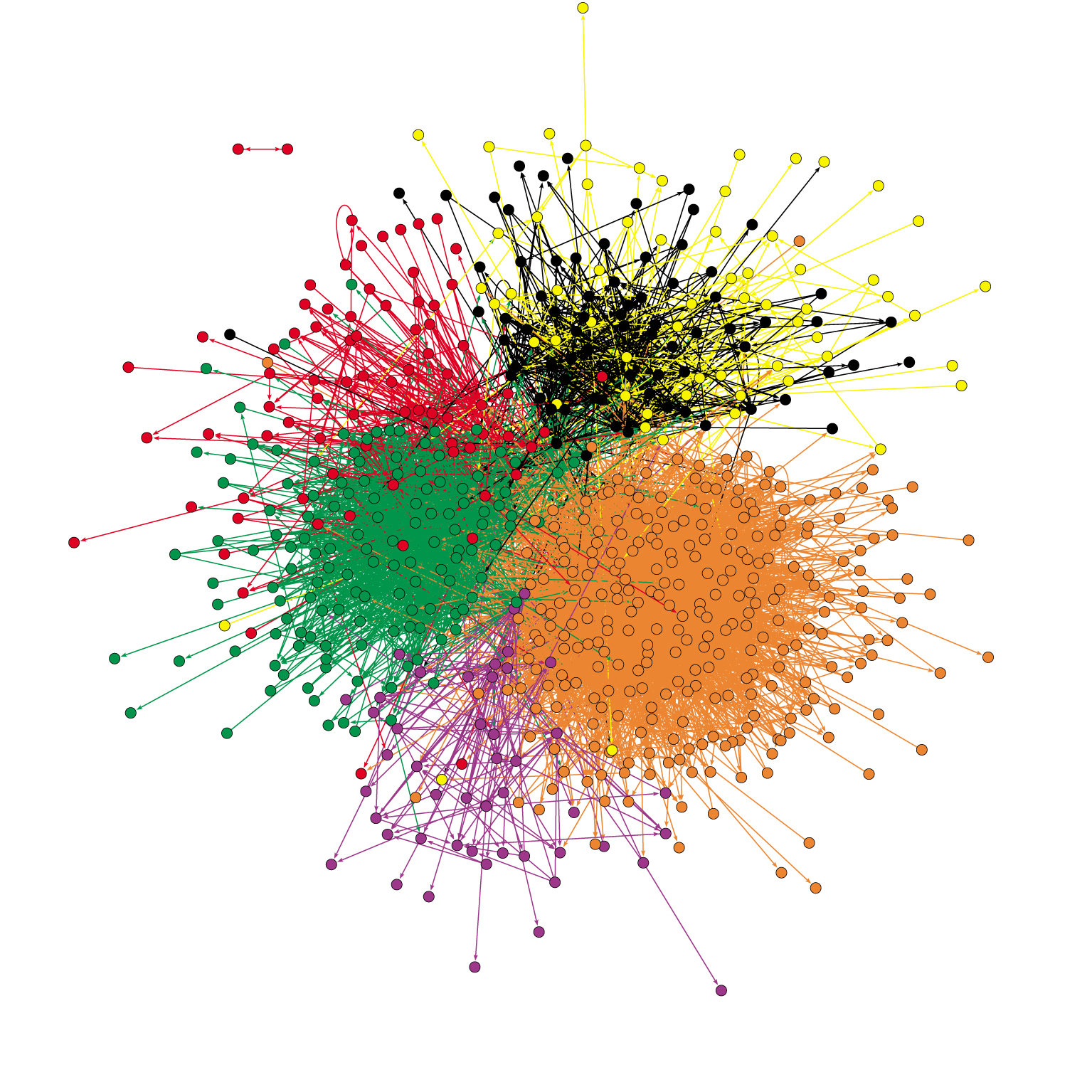}
			\label{retweetNetwork}}
			\subfigure[Mentioning ($H$=$0.79$)]{\includegraphics[width=0.30\textwidth]{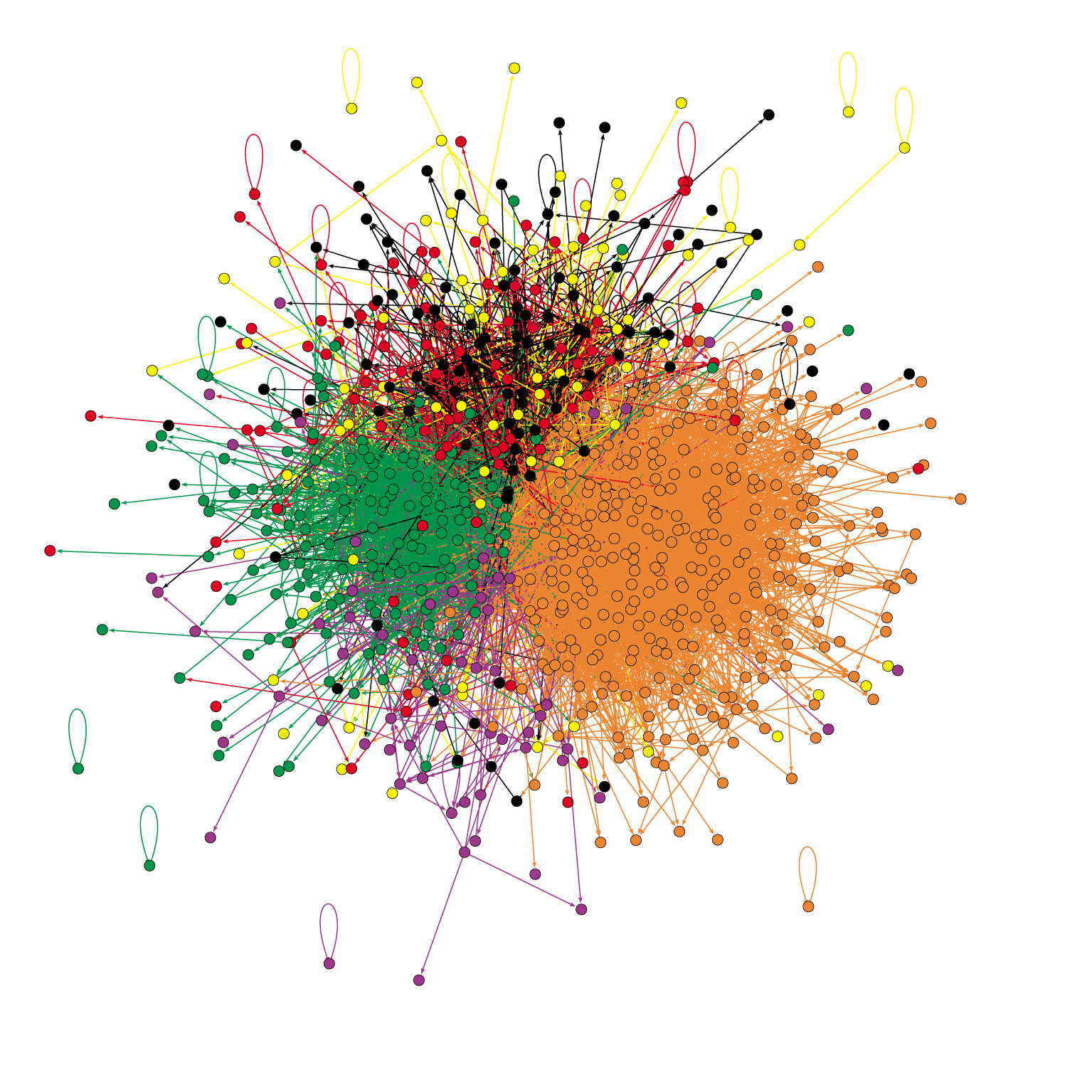}
			\label{mentionNetwork}}
		\end{tabular}
		\caption{\textbf{Examples of online conversational practices on Twitter}: Structures of the aggregate following, retweeting, and mentioning networks of German politicians from 9 weeks before to 4 weeks after the federal election 2013. The vertices in the networks correspond to user handles and are color-coded by party affiliation (colors given in Table \ref{statistics}). Arcs correspond to following/retweeting/mentioning relationships and are colored by sender. Structural differences between different practices can be observed: For example, homophily $H$ effects are lower in the mentioning network ($0.79$) than in the following ($0.83$) and retweeting ($0.90$) networks.
		CDU/CSU and FDP, which formed the last government coalition in Germany, are tightly knit in the follow and retweet networks.
		The Pirates are largely decoupled from a relatively pluralistic mentioning space where all other parties transact.
		The networks were laid out using the Kamada-Kawai algorithm.
		}
		\label{networks}
	\end{figure*}
	
	\begin{table*}[h!b!]
		\centering
		\caption{\textbf{Statistics and dataset for the German federal election 2013 on Twitter -- parties differ in several interesting ways:} Consistently across all conversational practices, the Pirate party exhibits the most homophilic behavior. Mentioning is most strongly used by the Pirates and the Greens ($D=0.07$), the two largest parties in terms of microblogging politicians (312 and 178), but not in terms of how many votes they actually received (2.2\% and 8.4\%). $D$ denotes network density, $\bar{k}$ denotes average in/out-degree of nodes, $\bar{w}$ denotes average in/out-weight of nodes, and $H$ denotes homophily, i.e., the tendency of a political party to communicate within party boundaries, computed on the individual level.}
		\begin{tabularx}{\textwidth}{l|c|c|c c c c|c c c c|c c c c} \hline
			& \bf{Election} & & \multicolumn{4}{ c| }{\bf{Following}} & \multicolumn{4}{ c| }{\bf{Retweeting}} & \multicolumn{4}{ c }{\bf{Mentioning}} \\
			\bf{Party} & \bf{Result} & \bf{Politicians} & \bf{\textit{D}} & \bf{\textit{\={k}\textsubscript{out}}} & \bf{\textit{\={k}\textsubscript{in}}} & \bf{\textit{H}} & \bf{\textit{D}} & \bf{\textit{\={w}\textsubscript{out}}} & \bf{\textit{\={w}\textsubscript{in}}} & \bf{\textit{H}} & \bf{\textit{D}} & \bf{\textit{\={w}\textsubscript{out}}} & \bf{\textit{\={w}\textsubscript{in}}} & \bf{\textit{H}} \\ \hline
			\cellcolor{black}\textcolor{white}{CDU/CSU} & 41.5\% & 158 & 0.15 & 32 & 38 & 0.78 & 0.08 & 14 & 14 & 0.86 & 0.04 & 16 & 22 & 0.64 \\ \hline
			\cellcolor{red}SPD & 25.7\% & 143 & 0.18 & 35 & 41 & 0.80 & 0.05 & 7 & 9 & 0.84 & 0.04 & 11 & 17 & 0.72 \\ \hline
			\cellcolor{yellow}FDP & 4.8\% & 143 & 0.17 & 35 & 35 & 0.78 & 0.05 & 7 & 9 & 0.84 & 0.03 & 6 & 9 & 0.55 \\ \hline
			\cellcolor{green}Greens & 8.4\% & 178 & 0.21 & 50 & 51 & 0.82 & 0.08 & 24 & 24 & 0.89 & 0.07 & 31 & 29 & 0.77 \\ \hline
			\cellcolor{purple}\textcolor{white}{Left} & 8.6\% & 97 & 0.23 & 30 & 32 & 0.79 & 0.07 & 8 & 13 & 0.91 & 0.04 & 13 & 14 & 0.68 \\ \hline
			\cellcolor{orange}Pirates & 2.2\% & 312 & 0.16 & 57 & 52 & 0.89 & 0.06 & 40 & 38 & 0.93 & 0.07 & 73 & 69 & 0.92 \\ \hline
			\bf{Total} & \bf{91.2\%} & \bf{1,031} & \bf{0.05} & \bf{44} & \bf{44} & \bf{0.83} & \bf{0.02} & \bf{25} & \bf{25} & \bf{0.90} & \bf{0.02} & \bf{39} & \bf{39} & \bf{0.79} \\ \hline
		\end{tabularx}
		\label{statistics}
	\end{table*}

	\section{Introduction}

	\label{sec:intro}

	In recent years, Twitter has established itself as a popular medium for public mass-personal communication \cite{Wu2011}.
	Our community has studied different aspects of Twitter communication, such as network structures (e.g., \citeauthor{Juergens2011} \citeyear{Juergens2011}; \citeauthor{Larsson2012} \citeyear{Larsson2012}), conversational practices (e.g., \citeauthor{Honeycutt2009} \citeyear{Honeycutt2009}; \citeauthor{Huang2010} \citeyear{Huang2010}), or analyses of dynamics (e.g., \citeauthor{Becker2011} \citeyear{Becker2011}; \citeauthor{2011ausvotes} \citeyear{2011ausvotes}).
	While this work has advanced our understanding about communication on Twitter, we know little about how these different perspectives can be integrated and quantified, and how they relate to conversational practices in a political context. Future computational social scientists would certainly benefit from computational tools and instruments that translate theoretical constructs from sociology to quantifiable measures that are amenable to computation and therefore can be applied on larger scales. 

	For example, it can be expected that in online conversations, political parties attempt to define themselves by adopting unique topical foci and conversational styles that are inherently difficult to assess. Having computational methods available to track and assess these styles would greatly increase our sociological understanding of such processes. In Figure \ref{networks}, we present preliminary evidence for this idea. The three networks depict the static outcome of communication among politicians on Twitter dissected according to politicians' following, retweeting, and mentioning practices. Table \ref{statistics} provides accompanying statistics about these networks. Interesting differences are present. For example, homophily effects are weakest for mentioning practices and the variance of homophily is also largest for this practice (cf. Table \ref{statistics}). This indicates that all parties tend to follow and retweet members of their own party but may use mentioning for inter-party debate. This finding is in line with previous results from a study on Twitter use before the U.S. midterm election of 2010 \cite{Truthy_icwsm2011politics}. Sound computational methods that would enable such qualitative insights into online conversational practices on a wide scale would represent a useful addition to the arsenal of social science methods.

	\textbf{Research questions:} These interesting yet preliminary differences in the static outcomes of conversational practices lead us to expect to see differences in the socio-cultural processes that produce these outcomes as well. While our observations above suggest that following, retweeting, and mentioning exhibit unique traits, we know little about the particularities and idiosyncrasies of these practices and their party-specific adoption.

	For example, what are the different purposes that tagging, retweeting, and mentioning serve in online political conversations? How do they differ from each other? How consistently are different practices used across different parties? Moreover, one would also expect that these different practices are effected to varying extents by external events or factors. For example, how would a (TV) debate or the day of the election itself effect the online conversational practices of parties in general or individual parties specifically?

	\textbf{Approach:} In this paper, we set out to answer these and related questions by presenting and applying a computational approach to assessing the socio-cultural dynamics of online conversational practices \emph{over time}.
	Our work is rooted in relational sociology, specifically in theoretical work that considers episodes of stability and change in practice \cite{Mohr2008,White2008,Fuhse2009,Padgett2012}.
	We are interested in making different aspects of online conversational practices, in particular cultural focus, - similarity, and - reproduction as well as institutions and punctuations, amenable to quantitative measurements.
	In doing so, we follow a \emph{deductive style of research}, deriving measures from a theoretical discussion of sociological constructs. While this enables us to root our measures in theory, it makes validation a challenging endeavour. To evaluate our approach nonetheless, we choose to apply it to a particular case, i.e., to conversational practices of political parties on Twitter before, during, and after the German federal election of September 22nd, 2013. This enables us to generate insights into the practical utility of our deductive approach in a real world scenario, as well as into the conversational practices of the case itself.

	\textbf{Contributions:} The contributions of our work are threefold:
	First, we present and discuss several sociological constructs related to conversational practices on a theoretical level.
	Second, we present a computational approach that deduces measures for each of the sociological constructs of interest. While the constructs are grounded in sociological theory, the proposed measures stem from computer science, social science, information science, and related fields.
	Third, to demonstrate the utility of our computational approach, we conduct a case study on the German federal election 2013 and present empirical insights into the conversational practices of German politicians during the course of this event.

	The paper is structured as follows: After related work we introduce the sociological background and constructs which form the theoretical foundation of our computational approach which we present subsequently. Finally, we describe our empirical study which demonstrates the utility of our approach, discuss our empirical insights, and conclude our work.

	\section{Related Work}

	\label{sec:relWork}
	Previous work which is relevant for our research focused either on analyzing the hashtagging, mentioning, and retweeting behavior of Twitter users in general or on analyzing the role of Twitter in a political context. To our best knowledge, ours is the first work which coherently studies the tagging, mentioning, and retweeting behavior of politicians as socio-cultural processes, and presents comparative analysis that yield unique insights into online conversational practices of politicians and political parties on Twitter.

	\smallskip
	\textbf{Hashtags, Mentions, and Retweets:} Twitter is used for many purposes, including the reporting of daily activities, communicating with other users, sharing information, and reporting, or commenting on, news \cite{Java2007}. As such it enables several \emph{conversational practices}.
	Hashtags are primarily used to describe news or communications with others and to find other users’ tweets about certain topics. Since tagging behavior is inspired by the observed use of hashtags in a users’ network \cite{Huang2010}, coherent semantic structures emerge from hashtag streams \cite{Wagner2010}. 
	Retweeting is the forwarding of other users’ tweets. By 2010, conventions as to how, why, and what users retweet had emerged, but the practice had not yet stabilized \cite{Boyd2010}. The retweetability of a Twitter message is related to its informational content and value and the embeddedness of its sender in following networks \cite{Suh2010}.
	Mentions, sometimes called @mentions or replies, emerged to be Twitter’s convention for the interactive use of addressing others, although it is also being used for other purposes like referencing. In 2009, most tweets without a mention reported daily activities while tweets with @ signs exhibited much higher variance in terms of topics and types of content \cite{Honeycutt2009}.

	\smallskip
	\textbf{Elections on Twitter:} With the rise of social media many researchers got interested in exploring the role of Twitter for politics. 
	Scientists from different backgrounds started exploring to what extent Twitter can predict election results with contradicting conclusions.
	Some research suggests that election results can be predicted by analyzing Twitter (e.g., \citeauthor{livne2011party} \citeyear{livne2011party}), while \citeauthor{Jungherr2013votes} (\citeyear{Jungherr2013votes}) shows that, at least for the German multi-party system, election result can not be predicted using Twitter. \citeauthor{MetaxasMG11} (\citeyear{MetaxasMG11}) conduct a meta study and conclude that electoral predictions using the published research methods on Twitter data are not better than chance.

	\textbf{Political discourse on Twitter:} Besides the interest in the voting behavior of persons and its reflection on Twitter, researchers also got interested in studying political discourse on Twitter. For example, \citeauthor{Truthy_icwsm2011politics} (\citeyear{Truthy_icwsm2011politics}) analyze the retweet and mention networks from 6 weeks leading up to the 2010 U.S. midterm election. Interestingly, the authors find extremely limited connectivity between right- and left-leaning users in the retweet network, but not in the mention network. This indicates that retweets and mentions are different conversational practices.
	The work of \citeauthor{Juergens2011} (\citeyear{Juergens2011}) shows that on Twitter new gatekeepers and ordinary users tend to filter political content based on their personal preferences. Therefore, political communication on Twitter might be highly dependent on a small number of users, critically positioned in the structure of the network.
	
	\textbf{Politicians on Twitter:} In previous research \citeauthor{thammbleier2013} (\citeyear{thammbleier2013}) found that retweets are motivated by professional uses while replies are used mainly for personal communication.
	\citeauthor{Schweitzer2011} (\citeyear{Schweitzer2011}) explored political e-campaigns in Germany and concluded that they increasingly reflect those patterns of traditional election coverage that have been held accountable for rising political alienation among the public, i.e., strategic news and extensive negativism.

	Following the suggestion of \citeauthor{Murthy2012} (\citeyear{Murthy2012}) who emphasizes the importance of a sociological understanding of Twitter, we now present an approach based on sociological constructs that allows to assess political online conversations.

	\section{Theoretical Constructs}
	\label{background}

	In this section, we elaborate the theoretical sociological constructs behind conversational practices which build the basis of our computational approach. 
	In doing so we start from sociological network theory known as relational sociology \cite{Carley1991,Mohr2008,White2008,Fuhse2009,Padgett2012}.
	The central premise is that social life is complex and stochastic and that \emph{identities}, which can be persons, groups, or higher-level agents, try to gain \emph{control} over the ensuing uncertainty through regularities. In many contexts, control is gained by collectively forming a densely clustered and culturally coherent community, triadic closure and homophily being the mechanisms \cite{mcpherson2001birds,Kossinets2009}.

	We first discuss an accessible understanding of culture and then offer an idealtype of how culture is reproduced by styles of practice. Because of the stochastic nature of social formations, we finally deal with system perturbation.

	\smallskip
	\textbf{Cultural Facts and Homophily:}
	Gaining control is a project in the socio-cultural space-time of simultaneous emergence of culture and feedback on structure: As agents act, they are not only engaged in transactions with other agents, they also make reference to pieces of information or \emph{facts} such as ideas, beliefs, concepts, symbols, knowledge, etc. From this practice, a culture emerges which can be understood as a distribution of referenced facts \cite{Carley1991}. But as agents in transactions are confronted with facts, existing culture positively or negatively feeds back onto agent behavior.
	Computer simulations are able to produce the densely clustered and culturally coherent communities which are ubiquitously found if cultural similarity breeds social connection \cite{Carley1991,Axelrod1997}.

	\smallskip
	\textbf{Styles, Institutions, and Reproduction:}
	If an identity has gained control and maintains it over time it has a \emph{style}. Styles are inert mechanisms of reproduction. It is due to the existence of such dynamical regularity that identities are predictable \cite{Kosinski2013,schoen2013power}. 
	Cycles of reproduction (styles) simultaneously determine practices of individuals and groups \cite{Mohr2008,White2008}.

	Distributions of facts like words, religious beliefs, surnames, and citations are fat-tailed \cite{Clauset2009}. \emph{Institutions} are cultural facts that are stably reproduced over time. Thereby, they become popular and are consequently found in the fat tail of probability distributions.
	Mechanisms of cumulative advantage \cite{Simon1955,Dellschaft2008,Papadopoulos2012} are theoretically compatible and have been shown to be capable of producing these skewed distributions.
	However, not all facts in the tail are necessarily institutions, since facts can also become popular due to short activity bursts rather than prolonged activity.

	\smallskip
	\textbf{Punctuations:}
	Continuity and normality is just one side of social life.
	Only in equilibrium persons and groups can reproduce uninterruptedly. In reality, social life is \emph{punctuated} by minor and major events that interrupt the normal flow of reproduction. Such punctuations may reflect scientific \cite{Kuhn1962}, political \cite{Brunk2001}, or economic \cite[Ch. 6]{Padgett2012} innovations or generally perturbations originating from inside or outside the observed system. Any account of socio-cultural processes must account for both normality and change \cite{Padgett2012}.
	It is both reproduction and punctuation that we study in this paper.

	\footnotesize
	\begin{table*}[h!b!]
	\centering
	\caption{\textbf{Operationalization of sociological constructs}.
	}
	\begin{tabularx}{\textwidth}{m{0.2\textwidth}|m{0.35\textwidth}|m{0.35\textwidth}}
	\hline
	\bf{Theoretical Construct} & \bf{Measure} & \bf{Description}  \\ \hline
	Cultural Focus $F$ & Shannon Entropy \cite{Shannon2001}& \emph{How strongly does a party focus on cultural facts?}  \\ \hline 
	Cultural Similarity $S$ & Cosine Similarity \cite{Baeza-Yates1999} & \emph{How similar are parties in terms of their culture vectors?} \\ \hline 
	Cultural Reproduction $R$ & Rank Biased Overlap \cite{Wagner2014}& \emph{How stable is a party's culture vector over time?} \\ \hline 
	Institutionness $I$ & Hirsch Index \cite{Hirsch2005} & \emph{How many weeks is a fact referenced that many times?} \\ \hline 
	Burstiness $B$ & Kleinberg's Burst Weight \cite{Kleinberg2003} & \emph{How popular is a fact in a given period relative to other facts?} \\ \hline 
	\end{tabularx}
	\label{operationalization}
	\end{table*}
	\normalsize

	\begin{figure*}[ht!]
	\centering

	\subfigure[Tagging Focus]{\includegraphics[width=0.23\textwidth]{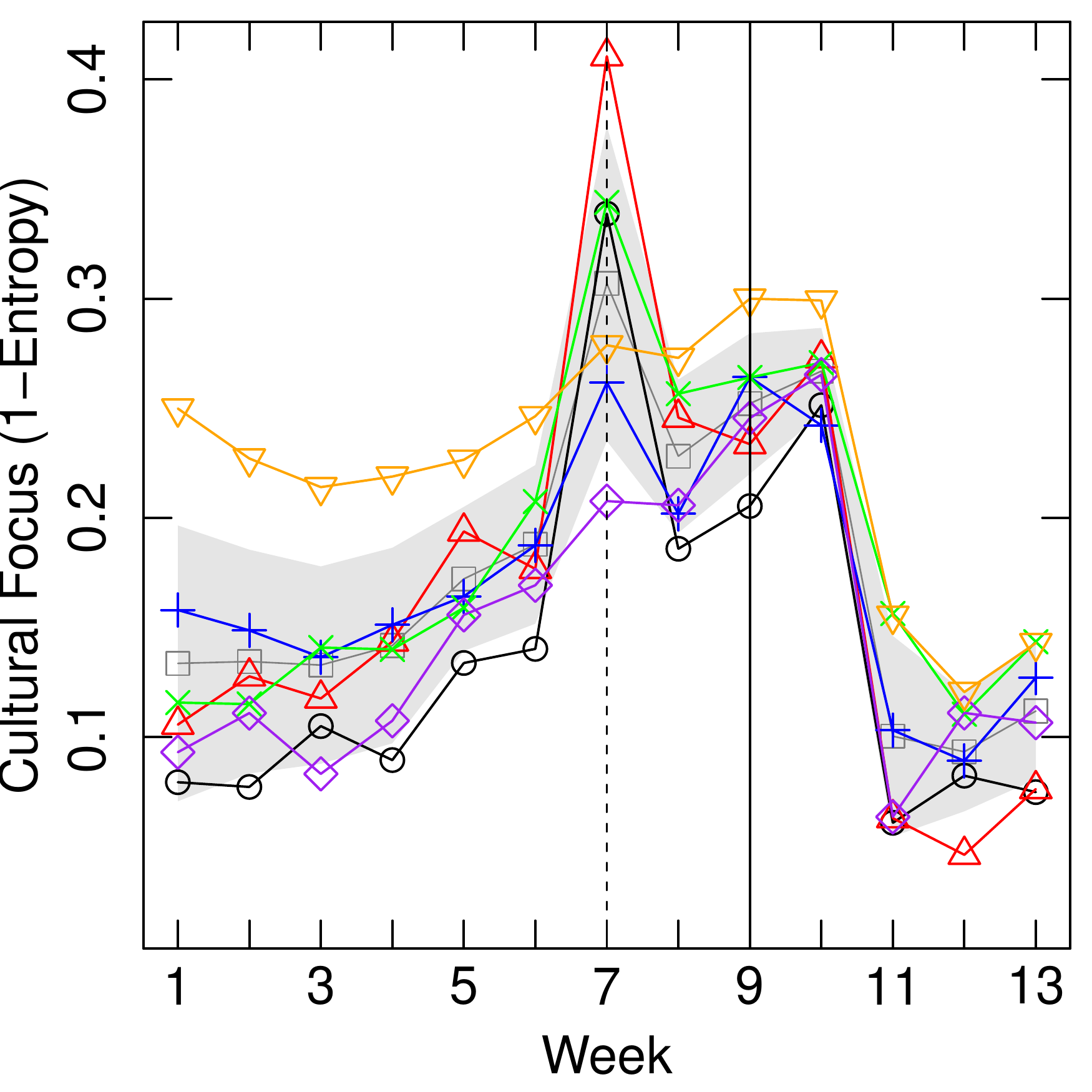}
	\label{TagEntr}}
	\subfigure[Retweeting Focus]{\includegraphics[width=0.23\textwidth]{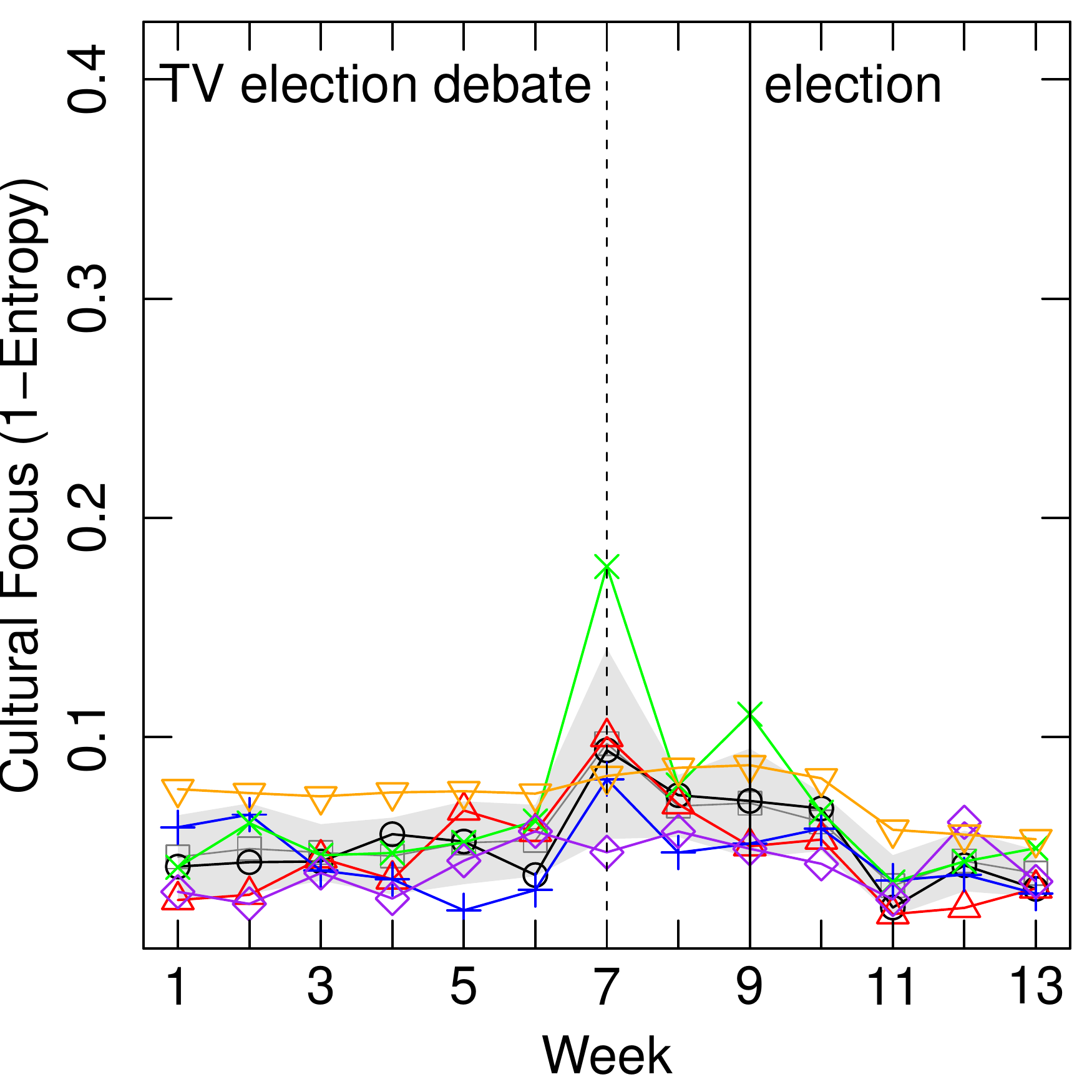}
	\label{RTEntr}}
	\subfigure[Mentioning Focus]{\includegraphics[width=0.23\textwidth]{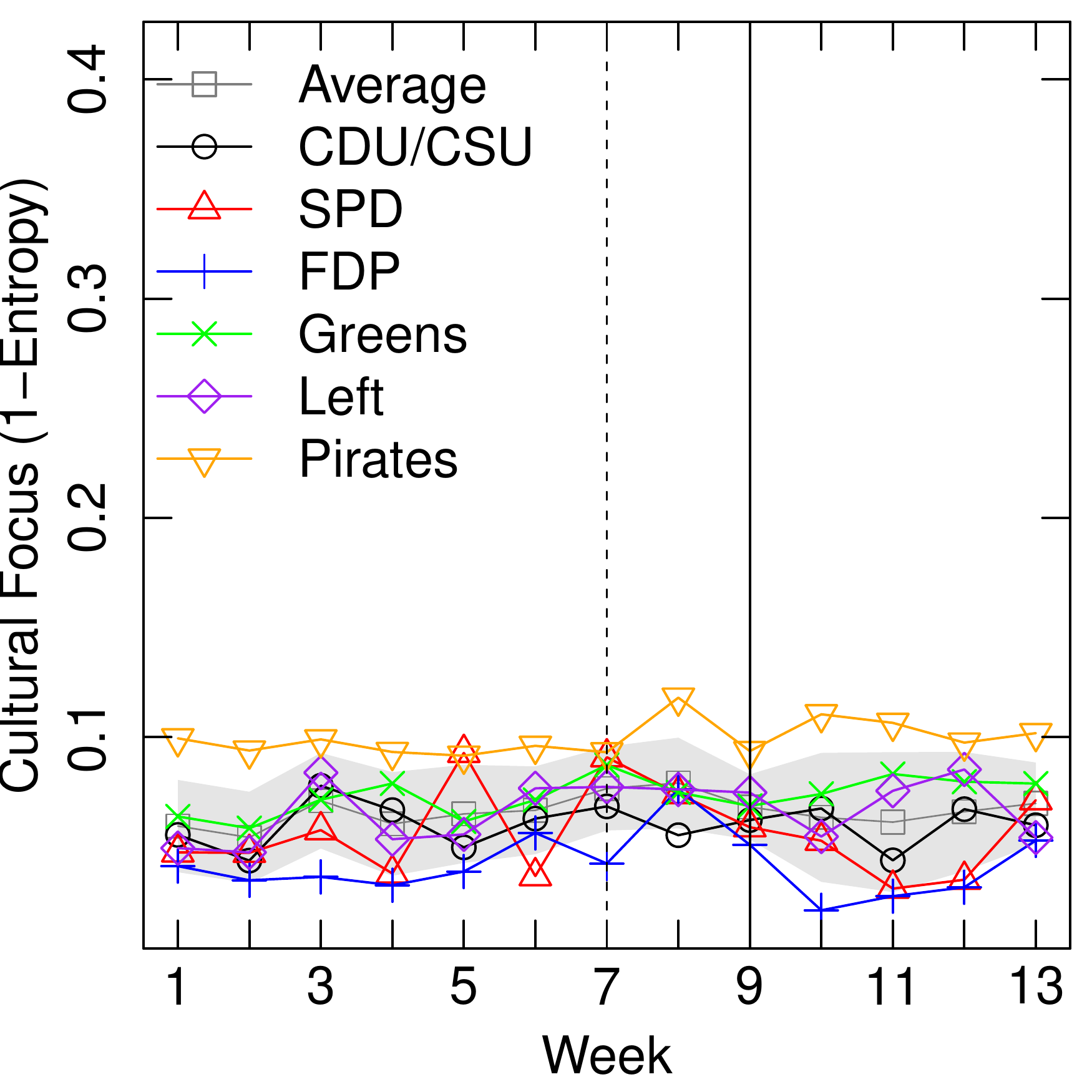}
	\label{@Entr}} \\

	\subfigure[Tagging Similarity]{\includegraphics[width=0.23\textwidth]{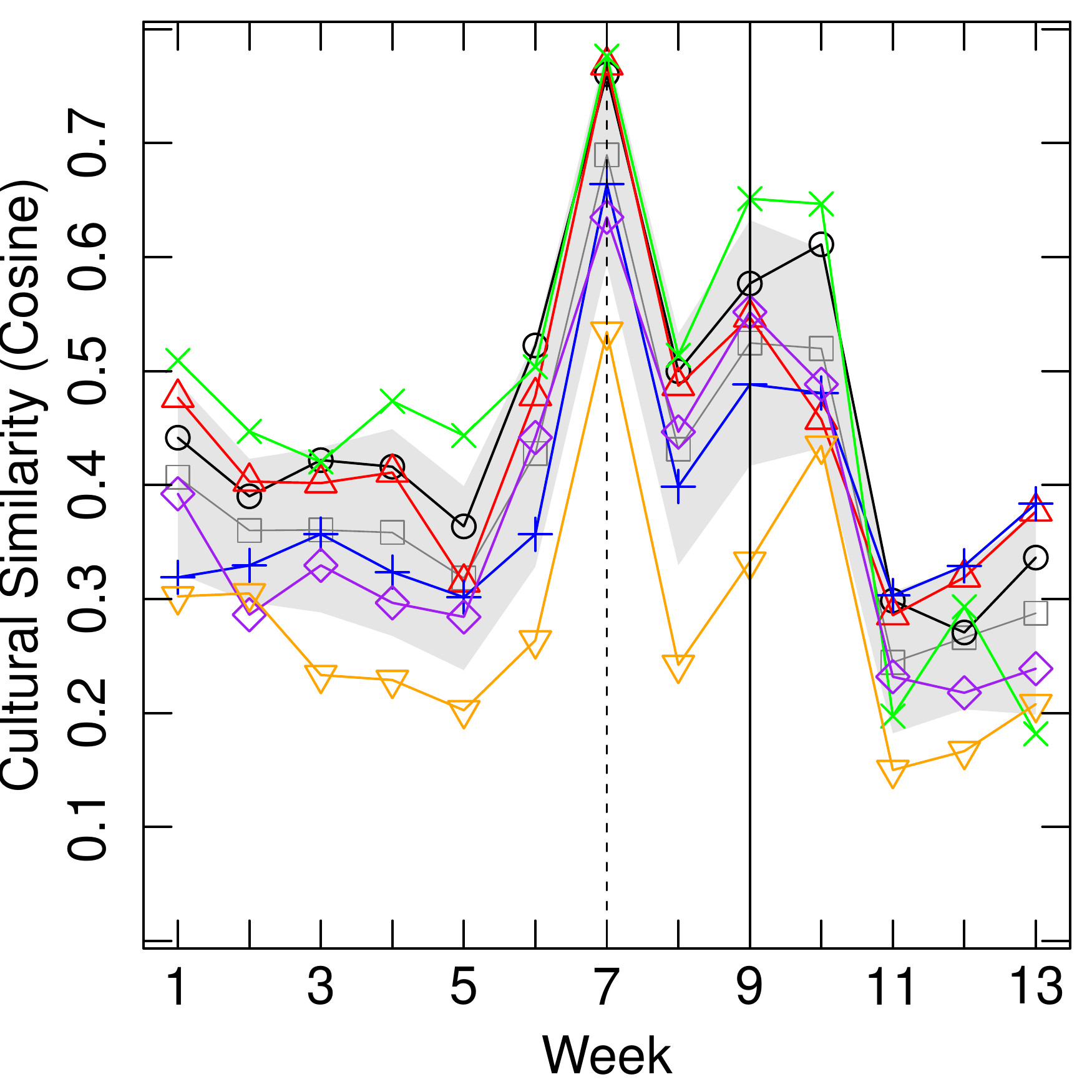}
	\label{TagCos}}
	\subfigure[Retweeting Similarity]{\includegraphics[width=0.23\textwidth]{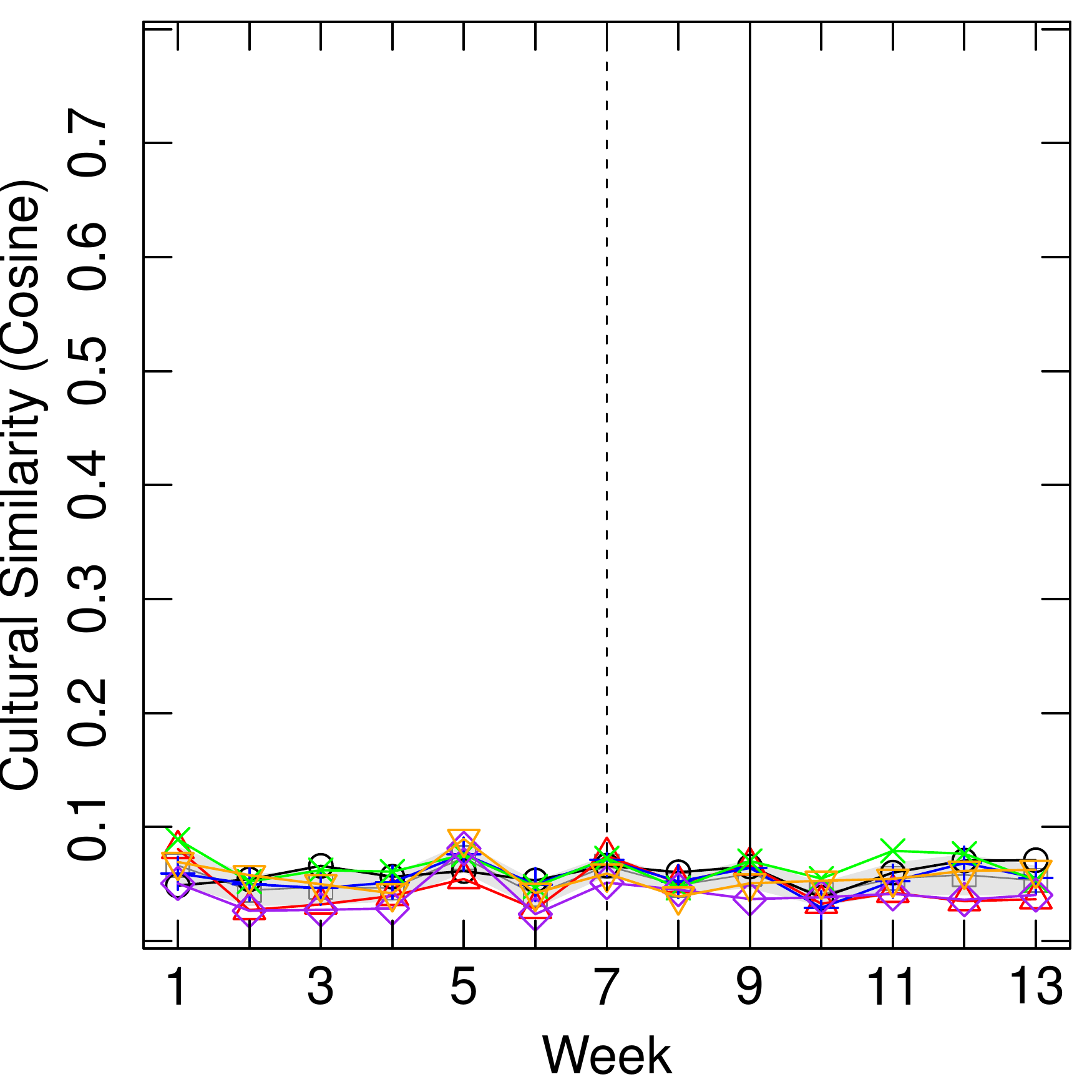}
	\label{RTCos}}
	\subfigure[Mentioning Similarity]{\includegraphics[width=0.23\textwidth]{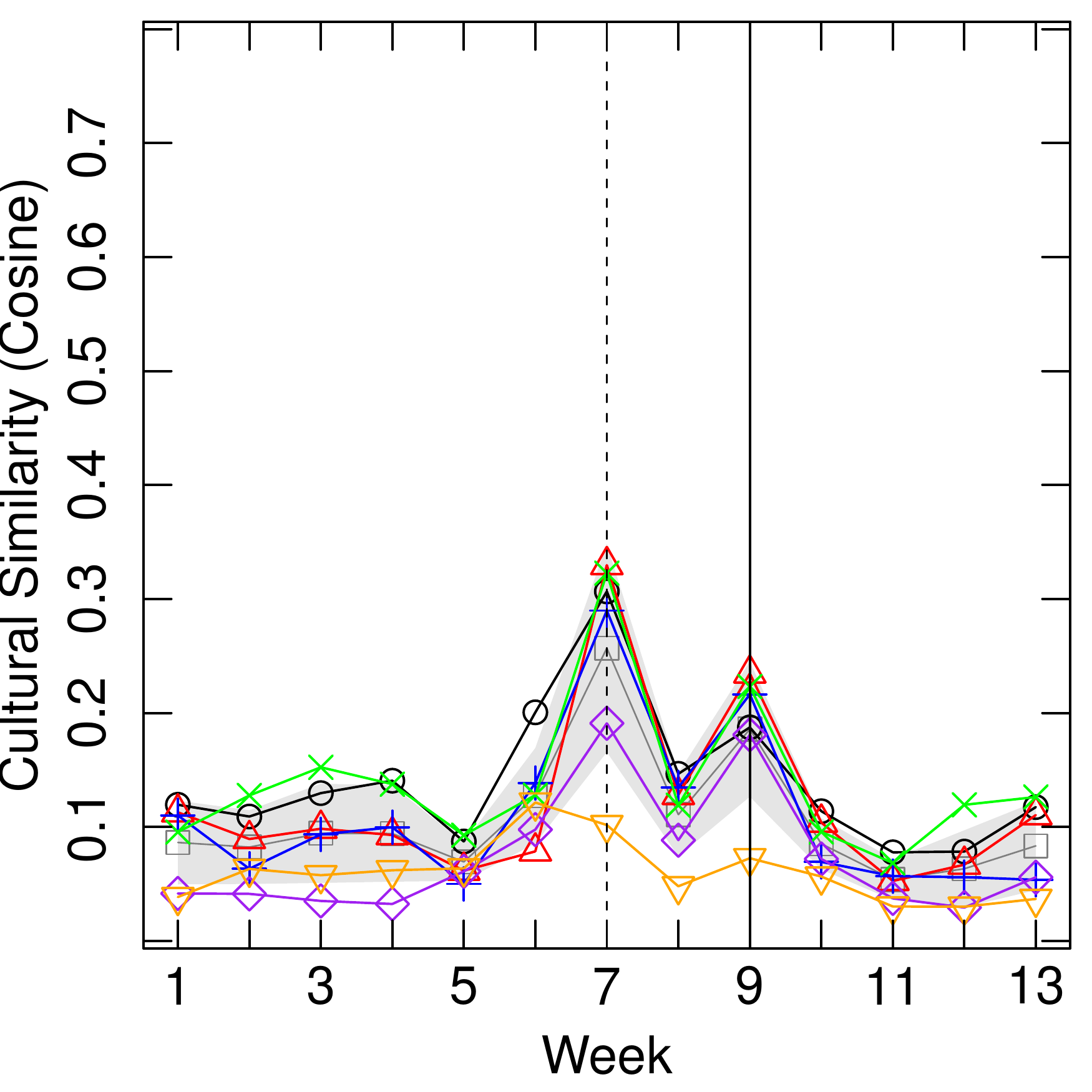}
	\label{@Cos}} \\

	\subfigure[Tagging Reproduction]{\includegraphics[width=0.23\textwidth]{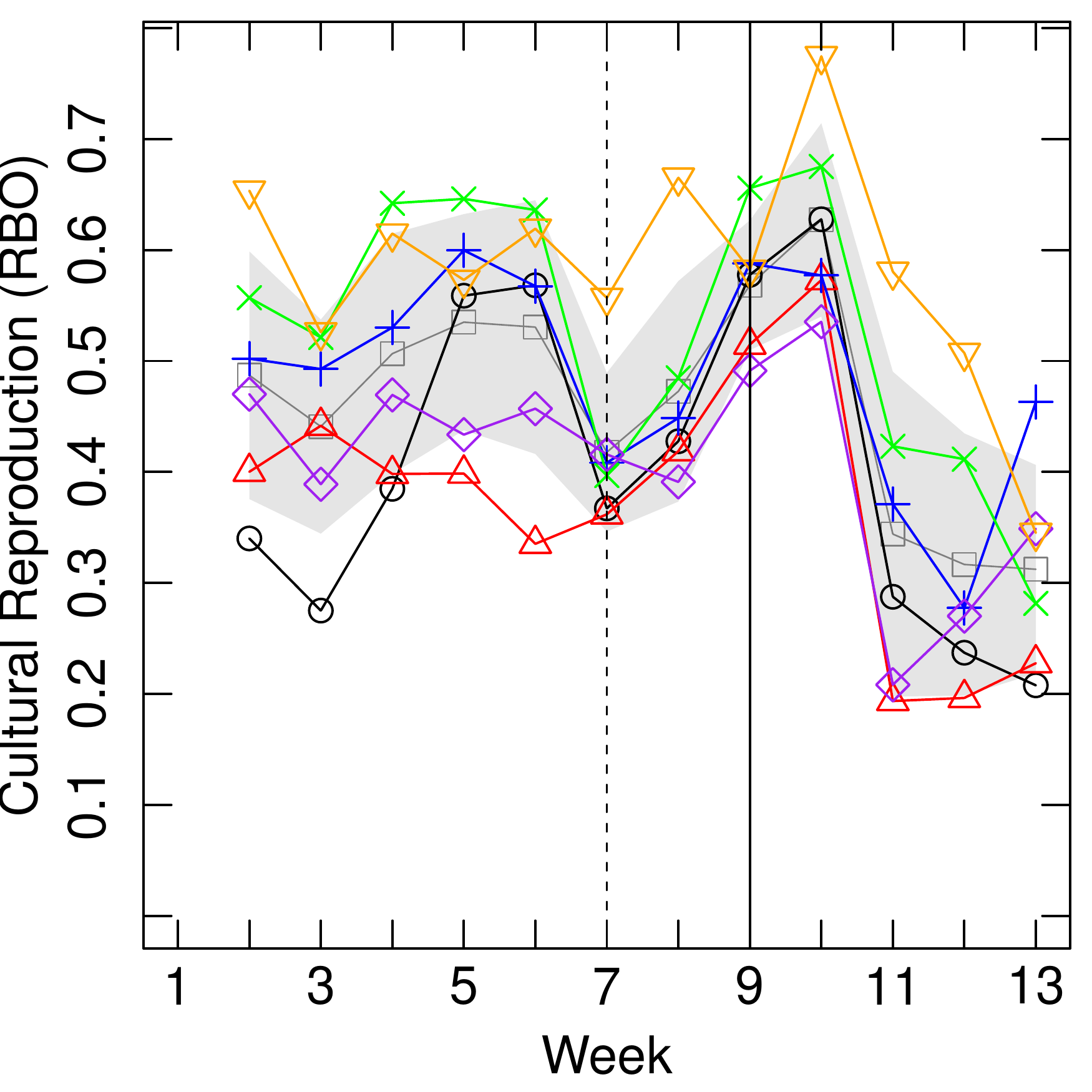}
	\label{TagStab}}
	\subfigure[Retweeting Reproduction]{\includegraphics[width=0.23\textwidth]{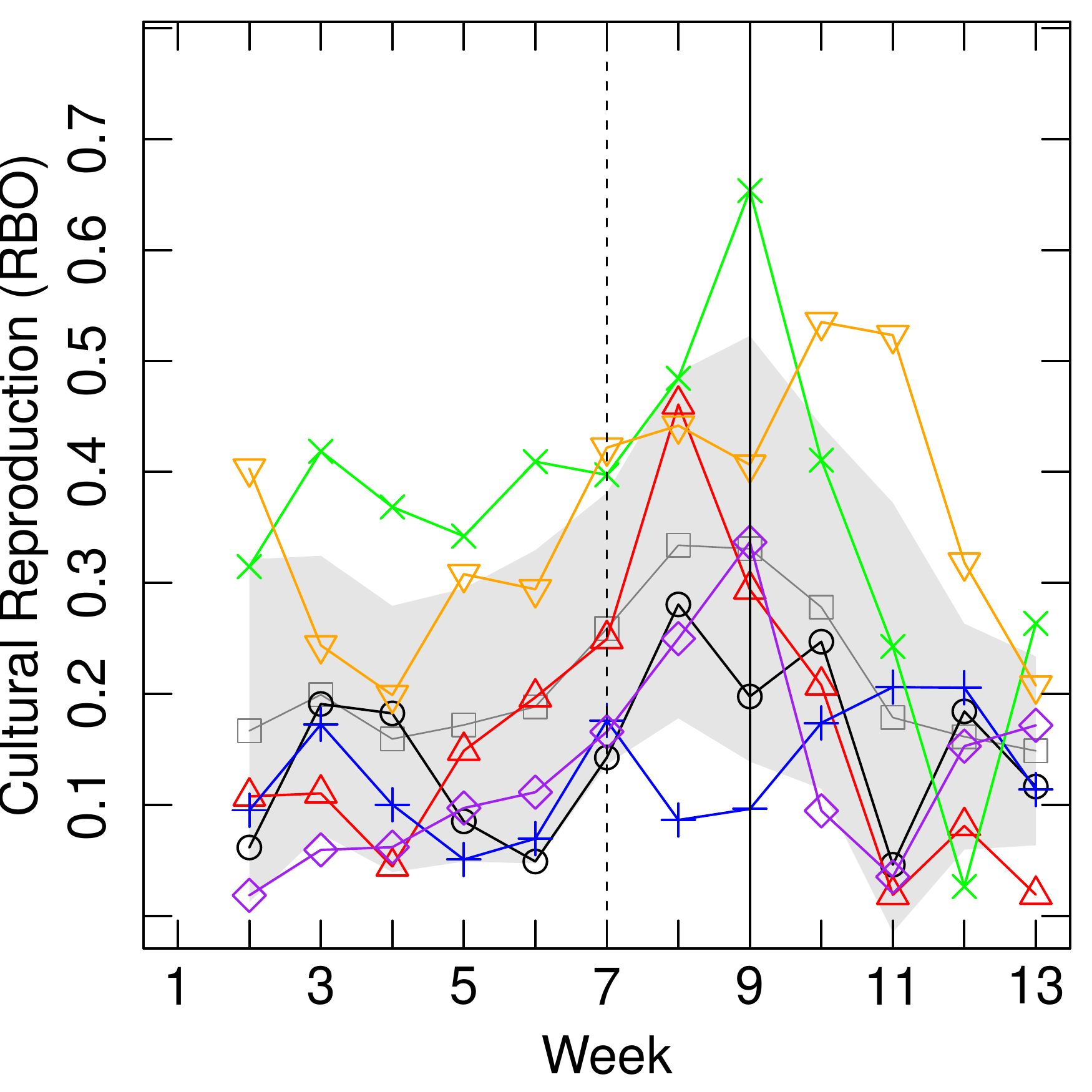}
	\label{RTStab}}
	\subfigure[Mentioning Reproduction]{\includegraphics[width=0.23\textwidth]{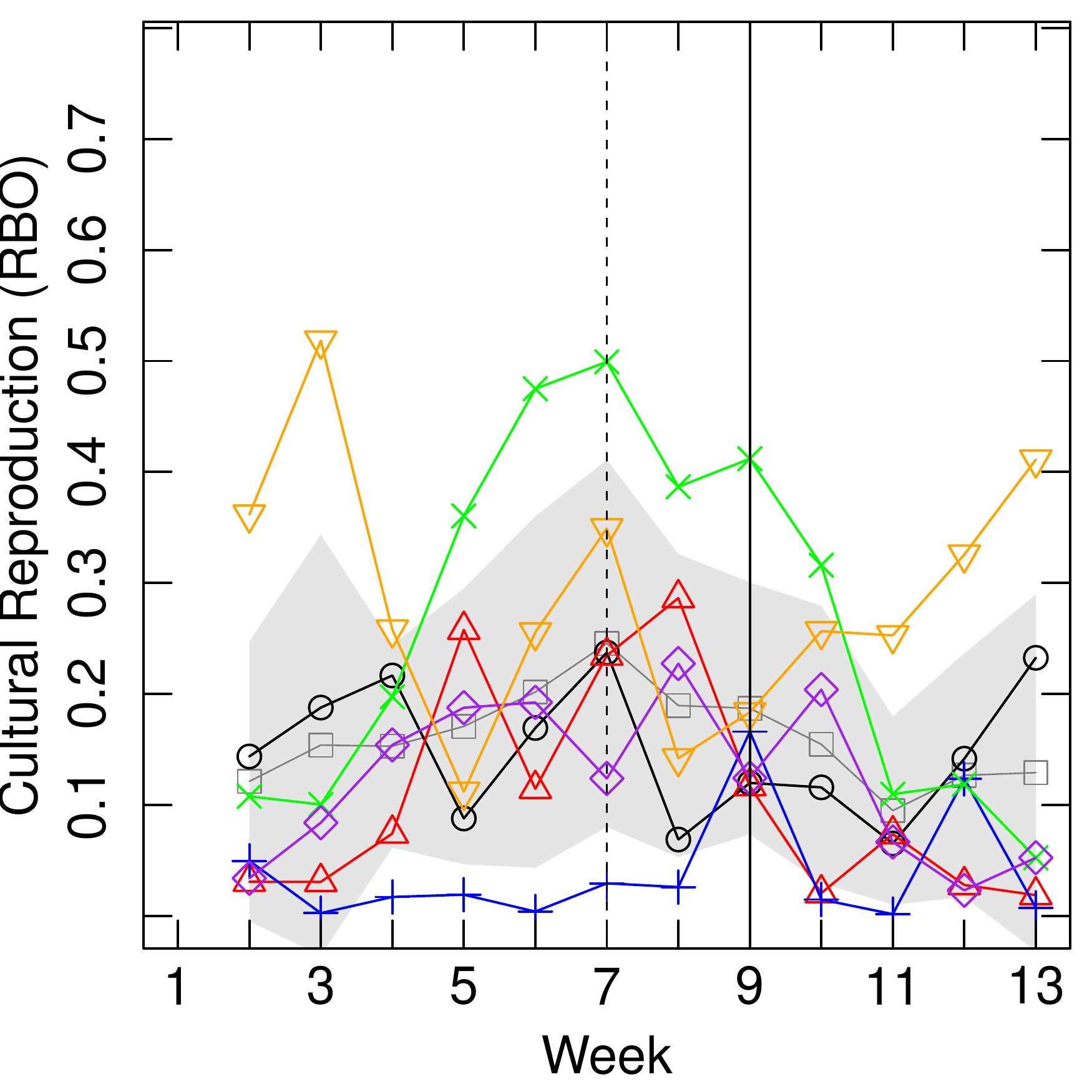}
	\label{@Stab}} \\

	\subfigure[Tagging Frequency]{\includegraphics[width=0.23\textwidth]{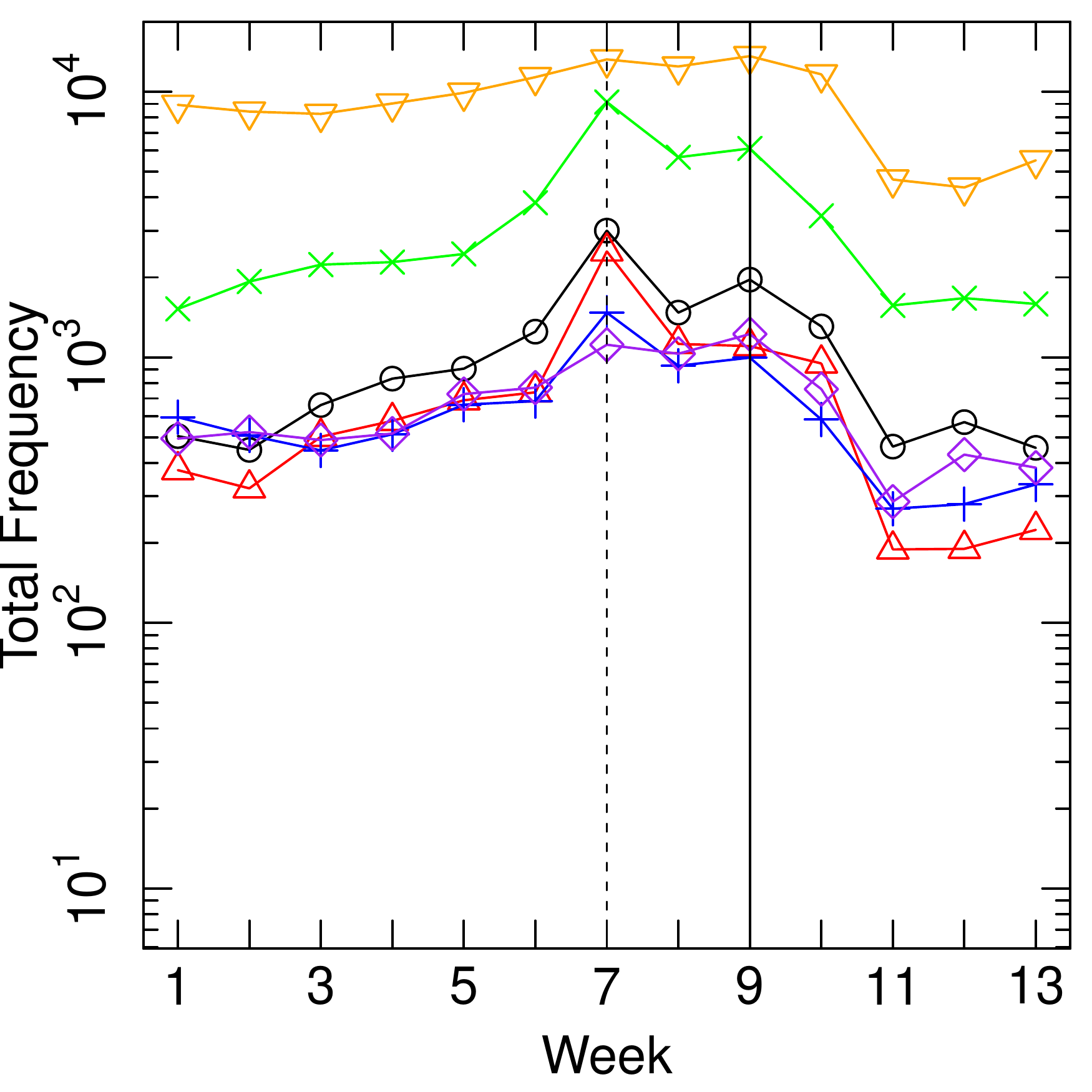}
	\label{TagFreq}}
	\subfigure[Retweeting Frequency]{\includegraphics[width=0.23\textwidth]{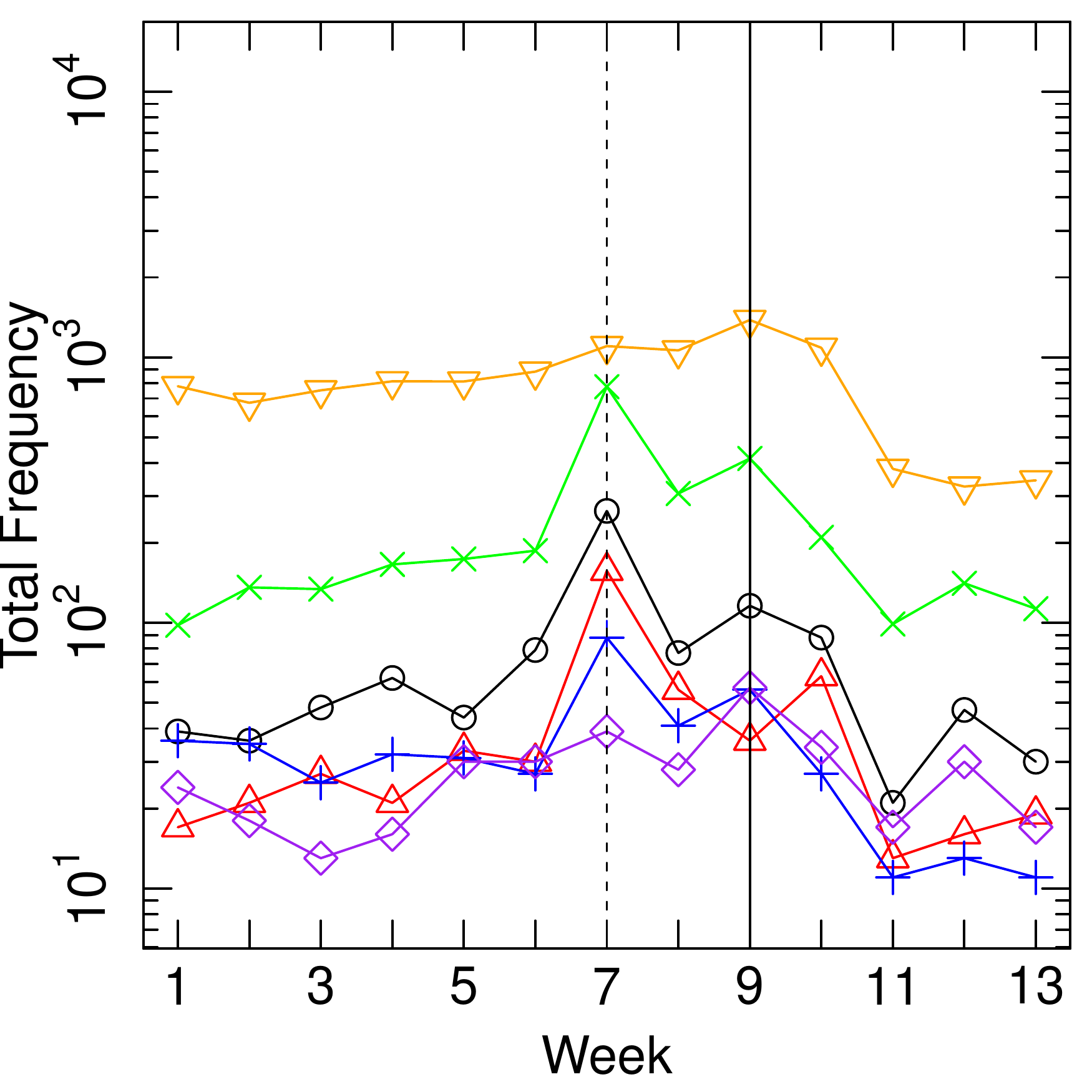}
	\label{RTFreq}}
	\subfigure[Mentioning Frequency]{\includegraphics[width=0.23\textwidth]{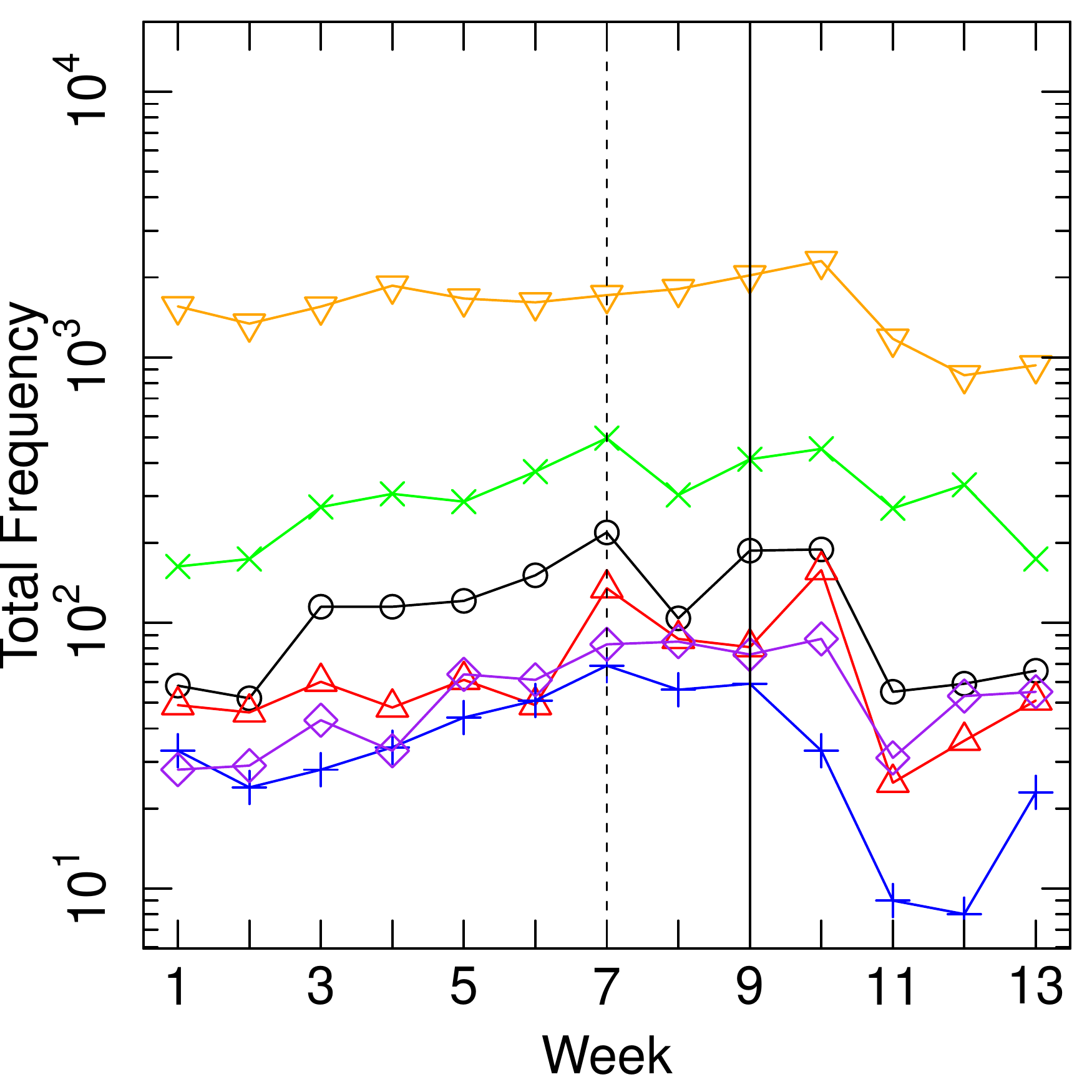}
	\label{@Freq}}

	\caption{\textbf{Online conversational practices of the six main political parties during the German federal election 2013:} Tagging, retweeting, and mentioning practices are shown over a period of 13 weeks. The three conversational practices (columns) are assessed using measures (rows) summed up in Table \ref{operationalization}.
	For the first three rows, curves for an average party are shown, obtained by averaging the six party scores. Shaded regions show standard deviations. The weeks of the election (straight line) and the TV election debate (broken line) clearly mark episodes of irregular activity. The last row shows the total number of references made to facts and is included as reference.}
	\label{dynamics}
	\end{figure*}
	
	\section{Operationalization of Constructs}
	\label{research_design}

	Next, we describe how we operationalize the sociological constructs described in the previous section for assessing online conversational practices of politicians and political parties on Twitter.

	\textbf{Approach:} The central idea behind conversational practices as socio-cultural processes is that a transaction -- a user using a hashtag, a user following another user, a user retweeting another user, or a user mentioning another user -- involves a social subject (the referencing user) and a cultural object (a referenced hashtag, followee, retweetee, or mentionee).
	In the case of following, retweeting, and mentioning, subjects and objects are of the same type, exemplifying the duality of the social and the cultural.
	In the empirical study which we will present later we are interested in how politicians practice online conversations on Twitter, concretely we study their mentioning, retweeting, and hashtagging practices. We use aggregations of users (parties) as \emph{objects of study} and referenced facts (i.e., user handles and hashtags) as \emph{units of analysis}.
	For analyzing dynamics, tweets are batched into bins of one week each.
	Though our operationalization is specific to the context of our empirical study (politics on Twitter), our computational approach for assessing online conversational practices is general and can be applied to study the practices of other groups of users on Twitter (or even on other stream-based social systems which provide support for conversations and allow to observe the referencing of facts over time).

	\subsection{Cultural Focus and Similarities}

	In the following, we interpret the observable cultural objects (i.e., hashtags and user handles) of conversational practices (i.e., retweeting, hashtagging, and mentioning) of Twitter users as cultural \emph{facts}.
	The culture of a group of users (which corresponds to a party $i$ in our case) is represented by a vector $\sigma_{i}=(a_1, a_2, ..., a_n)$ where the elements are the frequencies of (a) the hashtags that members of party $i$ used, (b) the users that members of party $i$ retweeted, or (c) the users which members of party $i$ mentioned within the observation period.

	\smallskip
	\textbf{Cultural Focus:}
	To quantify the extent to which a party $i$ reveals a \emph{cultural focus} $F$ on few selected hashtags or users (facts), the normalized Shannon entropy \cite{Shannon2001} -- a measure of disorder -- of the party's fact vector is used:
	\begin{equation}
	F(\sigma_i) = 1-\frac{-\sum_{j=1}^{n} p(a_j)*\log_2 p_(a_j)}{\log_2(n)}
	\end{equation}
	Here, $p(a_j)$ corresponds to the frequency of a cultural fact $a_j$ for party $i$ divided by the frequency of all other facts of that party.
	Because the denominator holds the maximum entropy and normalized entropy is subtracted from 1, $F$ falls in the range $[0,1]$ where 0 means that there is no focus (highest entropy) and 1 that the focus is highest (no entropy).

	\smallskip
	\textbf{Cultural Similarity:}
	Culturally similar agents are more probable to interact with each other. To reveal potential homophily effects, cultural similarities of parties are studied over time.
	We propose to measure the \emph{cultural similarity} $S$ of two parties $i$ and $j$ using the cosine similarity \cite{Baeza-Yates1999} of their fact vectors:
	\begin{equation}
	S(\sigma_i, \sigma_j) = \frac{\sigma_{i} \cdot \sigma_{j}}{\parallel \sigma_{i}\parallel \parallel \sigma_{j} \parallel}
	\end{equation}

	Because facts cannot have negative frequencies, similarities are in the range $[0,1]$ where 0 indicates no similarity and 1 highest similarity. Cosine measures the similarity for individual party pairs. To obtain a score for a single party $i$, we compute its average cosine similarity to all other parties $j$. We consciously do not weight scores by user size or tweet volume because we operate on the level of parties as objects of study. 

	\subsection{Styles, Institutions and Reproduction}
	\smallskip
	\textbf{Cultural Reproduction:}
	Styles are mechanisms of reproduction of focus.
	To study and quantify styles, we proceed in two steps. First, we measure the reproduction of cultural focus from week to week. In a second step, the stability of reproduction over time is studied to generate a quantitative judgement about styles.
	Reproduction is operationalized by an extended version of the Rank Biased Overlap (RBO) metric \cite{Wagner2014} which allows to measure stability in social streams. The \emph{cultural reproduction} $R$ of a party $i$ is defined as follows:
	\begin{equation}
	R_{i}(\sigma_{i}, p) = (1-p) \sum_{d=1}^{\infty} \frac{2 \cdot \sigma_{i}^{1:d}(t1) \cap \sigma_{i}^{1:d}(t2) }{|\sigma_{i}^{1:d}(t1) + \sigma_{i}^{1:d}(t2)|} p^{(d-1)}
	\end{equation}
	The cultural fact vectors $\sigma_{i}(t1)$ and $\sigma_{i}(t2)$ of party $i$ hold the frequencies of the facts which have been observed
	at $t1$ and $t2$, respectively.
	These vectors are not necessarily conjoint since new facts can be introduced at any point in time.
	Let $\sigma_{i}^{1:d}(t1)$ and $\sigma_{i}^{1:d}(t2)$ be the ranked vectors at depth $d$.
	The scores fall in the range $[0,1]$ where 0 means that the culture of a party $i$ at $t1$ is completely different from its culture at $t2$ (culture is unstable, turnover is maximum), and 1 means that it is identical (i.e., does not change over time and is therefore perfectly stable).
	The parameter $p$ ($0 \leq p < 1$) determines how steep the decline in weights is.
	The smaller $p$ is, the more tail-weighted the metric is, the tail being the high frequency part of the distribution.
	We chose $p=0.9$ which means that facts in the fat tail are most relevant for assessing the cultural stability of a party compared to changes in the head of the distribution (86\% of the weight is given to the 10 highest ranked cultural facts if $p=0.9$).
	This makes intuitively sense because, if a party changes a cultural fact which is core to them, they will become very unstable according to our measures, while, if they change a fact which is just one among many with low frequencies, this will only have a small impact on their stability.

	Mathematically, a style is present when $\delta R_{i}/\delta t \geq 0$. If $R$ decreases over time, the party will dissolve into chaos as less and less facts are reproduced. If $R$ increases, the party will eventually freeze as reproduction reaches 1 and cultural facts will forever be the same. From a systems perspective, a constant turnover of facts is desired.

	\smallskip
	\textbf{Institutionness:}
	Cultural reproduction is strongly related to the construct of stable facts called institutions which are expected to be found in the fat tail of the probability distribution of cultural facts. 
	Two notions resonate with the idea of stable facts: that they are (a) highly referenced and (b) continuously highly referenced.
	It seems straightforward to divert the Hirsch index \cite{Hirsch2005} from its intended use.
	Originally proposed to quantify an individual's scientific research output, a scientist has an index $h$ if $h$ of his papers have at least $h$ citations each, and all but $h$ papers have no more than $h$ citations each.
	Applied to our temporal case, a fact has an index $h$ if for $h$ weeks it is referenced at least $h$ times, and in all but $h$ weeks no more than $h$ times.
	A normalization -- one part of which is a division of a paper's citation rate by the citation rate of an average paper in the subject area -- has been proposed to make the Hirsch index comparable across scientific disciplines with different publication and citation characteristics \cite{radicchi2008universality}.
	In analogy, a fact $a$ has index $h$ if for $h$ weeks it is referenced at least $h/h_{0}$ times, and in all but $h$ weeks no more than $h/h_{0}$ times, where $h_{0}$ is the week-specific reference rate of an average fact.
	Let's call this index the \emph{institutionness} $I$ of fact $a$.
	Since data ranges over 13 weeks, $I$ is in the range $[0,13]$.
	
	\subsection{Punctuations}

	\textbf{Burstiness:}
	Having a style means being predictable. Punctuations interrupting the normal flow of reproduction leave traces in Twitter in the form of activity bursts.
	To operationalize, we refer to \citeauthor{Kleinberg2003}'s (\citeyear{Kleinberg2003}) observation that social streams are ``punctuated by the sharp and sudden onset of particular episodes''.
	A fact is considered to burst if it leaves a period where its popularity relative to other facts is small and enters a period in which its relative popularity is reasonably large.

	Suppose there are $n$ batches of transactions made by party $i$ (in our case $n=13$ since we use weeks as batches). The $t$-th batch contains $r_t$ transactions with references to fact $a$ 
	out of a total of $d_t$ transactions.
	Let $R=\sum_{t=1}^{n}r_t$ and $D=\sum_{t=1}^{n}d_t$ and $p_{s}=\frac{R}{D}2^{s}$.
	To reveal the burstiness of a fact $a$, we compute for each week $t$ the party-specific cost of that fact to be transformed in a burst state $s$:
	\begin{equation}
		\gamma(s, r_t, d_t) = -\ln\left(\binom{d_t}{r_t}p_{s}^{r_t}(1-p_{s})^{d_t-r_t}\right)
	\end{equation}
	The weight of a fact's burst -- its \emph{burstiness} $B$ -- in the period $[t1,t2]$ is the improvement in cost by being in state $s=1$ for the period $[t1,t2]$:
	\begin{equation}
		B(a,t1,t2) = \sum_{t=t1}^{t2}\left(\gamma(0, r_t, d_t)-\gamma(1, r_t, d_t)\right)
	\end{equation}

	In words, for each bursting fact $a$ its burstiness in a period $t1$ (onset) to $t2$ is obtained. Facts can burst multiple times.
	There is no upper bound for a fact's burst weight. The final burstiness measure is normalized by the weight of a party's strongest burst and is, therefore, also in the range $[0,1]$.

	\tiny
	\begin{table*}[h!b!]
		\caption{\textbf{Institutionness $I$ and burstiness $B$ of selected facts:} The election received continuous attention by all parties (high $I$ scores) but bursted only for some parties (it was the Left's strongest burst). The short-lived TV debate, on the other hand, bursted strongly for all parties. In the process, the one debater's hashtag \#merkel and the other debater's username @peersteinbrueck also bursted for almost all parties. Square brackets give onset and end of burst episodes.}
		\tabcolsep=0.11cm

		\begin{tabularx}{\textwidth}{m{2.7cm}|m{3.5cm}|>{\centering\arraybackslash}m{0.4cm}>{\centering\arraybackslash}m{1cm}|>{\centering\arraybackslash}m{0.4cm}>{\centering\arraybackslash}m{1cm}|>{\centering\arraybackslash}m{0.4cm}>{\centering\arraybackslash}m{1cm}|>{\centering\arraybackslash}m{0.4cm}>{\centering\arraybackslash}m{1cm}|>{\centering\arraybackslash}m{0.4cm}>{\centering\arraybackslash}m{1cm}|>{\centering\arraybackslash}m{0.4cm}>{\centering\arraybackslash}m{1cm}} \hline

			& & \multicolumn{2}{c| }{{\bf{CDU/CSU}}} & \multicolumn{2}{c| }{\bf{SPD}} & \multicolumn{2}{c| }{\bf{FDP}} & \multicolumn{2}{c| }{\bf{Greens}} & \multicolumn{2}{c| }{{\bf{Left}}} & \multicolumn{2}{c }{\bf{Pirates}} \\

			\bf{Fact} & \bf{Description} & \bf{\textit{I}} & \bf{\textit{B}} & \bf{\textit{I}} & \bf{\textit{B}} & \bf{\textit{I}} & \bf{\textit{B}} & \bf{\textit{I}} & \bf{\textit{B}} & \bf{\textit{I}} & \bf{\textit{B}} & \bf{\textit{I}} & \bf{\textit{B}} \\ \hline
			\#btw13 & Federal election (Sep 22) & 13 & 0 & 11 & 0 & 10 & 0.69 [8,10] & 13 & 0.18 [10,10] & 11 & 1.00 [8,10] & 13 & 0.46 [10,10] \\ \hline
			\#tvduell & TV election debate (Sep 1) & 3 & 1.00 [7,7] & 3 & 0.94 [7,7] & 1 & 1.00 [7,7] & 3 & 0.66 [7,7] & 1 & 0.75 [7,7] & 2 & 0.85 [7,7] \\ \hline
			\#merkel & CDU/CSU, chancellor & 11 & 0.10 [7,7] & 11 & 0.25 [7,7] & 4 & 0.25 [7,7] & 13 & 0.18 [7,7] & 8 & 0.39 [6,7] & 13 & 0.10 [7,7] \\ \hline
			RT sigmargabriel & SPD, chairman & 1 & 0 & 3 & 1.00 [7,8] & 0 & 0 & 2 & 0.26 [5,8] & 0 & 0 & 0 & 0 \\ \hline
			RT Volker\_Beck & Greens, parliamentary managing director & 2 & 0 & 3 & 0 & 1 & 0 & 8 & 1.00 [7,7] & 1 & 0 & 1 & 0.26 [6,7] \\ \hline
			@peersteinbrueck & SPD, candidate for chancellor & 5 & 1.00 [7,8] & 8 & 1.00 [6,8] & 5 & 0 & 6 & 0.78 [7,8] & 2 & 0 & 2 & 0.24 [7,7] \\ \hline
		\end{tabularx}
		\label{facts}
	\end{table*}
	\normalsize
	
	\section{The German Federal Election 2013}

	In this section we describe the dataset which we used to conduct our empirical study of three conversational practices of politicians before, during, and after the German federal election 2013. Then, we present our results. In particular, we present our findings on hashtagging, mentioning, and retweeting practices of German politicians during the 2013 elections. While our approach would also be applicable to analyze following practices (as introduced in Figure \ref{hashtagNetwork}), we could not perform this analysis since the fine-grained temporal data necessary to study the evolution of follower networks on Twitter is impossible to collect via the public Twitter API.

	\subsection{Dataset}

	The dataset consists of German politicians and their communicative transactions on the microblogging platform Twitter between July 20th and October 18th, 2013. This corresponds to about 9 weeks before and 4 weeks after the federal election of September 22nd. A preliminary version of the ``Twitter corpus of candidates'' described by \citeauthor{Kaczmirek2013} (\citeyear{Kaczmirek2013}) is used. Our dataset contains 1,031 user account handles, 98\% of which were the ``most relevant'' candidates for the German parliament in early 2013, and their tweets. 
	983 politicians were active in the sense that they either tagged a tweet or followed, retweeted, or mentioned another politician during our window of observation.
	In total,
	there are 123,819 tweets containing at least one hashtag, 16,292 tweets retweeting at least one other politician (identified by ``RT username''), and 25,778 tweets mentioning at least one other politician (identified by ``@ username'').

	Each politician belongs to one political party. The Social Democratic party SPD is the oldest (150yrs), followed by the Conservatives CDU/CSU (68yrs), the Liberals FDP (65yrs), the Greens (Die Grünen, 34yrs), the Left (Die Linke, 24yrs), and the new, internet-affine Pirates (Piratenpartei, 7yrs).

	\subsection{Results}

	Before going into the details of the dynamics that created the networks shown in Figure \ref{retweetNetwork} and \ref{mentionNetwork}, we first address the two episodes that left spikes in the curves of Figure \ref{dynamics} which visualizes the main dynamics results.
	Both relate to offline events.
	The first is the election of September 22nd in week 10. Message volumes, and average tagging focus and retweeting reproduction rate increase towards the events and decrease afterwards. 
	Though the election day was early in week 10, we can see that most of the communication occurred in week 9. Therefore, the horizontal line marking the event in Figure \ref{dynamics} is in week 9.
	The second is the TV election debate between the two candidates for chancellor, A. Merkel (CDU) and P. Steinbrück (SPD), of September 1st, early in week 7.
	One can tell because the corresponding hashtag \#tvduell is the strongest burst for the CDU/CSU and FDP, the second strongest for the SPD and the Greens, the third strongest for the Left, and the fourth strongest for the Pirates, restricted to week 7 in each case.
	See Table \ref{facts} for the burstiness of selected facts.
	Despite the election being the main event in this stream, the debate marks the maximum in terms of total message frequency (cf. Figure \ref{TagFreq}) and the average tagging focus and similarity (cf. Figure \ref{TagEntr} and \ref{TagCos}). It also left spikes in the average retweeting focus and mentioning similarity (cf. Figure \ref{RTEntr} and \ref{@Cos}).
	In the following, we will discuss these events, average party dynamics, as well as the peculiarities of single parties.

	\smallskip
	\textbf{Cultural Focus:}
	The two major events in the stream have different impacts on the three practices. Figure \ref{TagEntr} shows that all parties almost monotonously concentrate their focus on hashtags towards the election (average increases from 0.13 in week 1 to 0.27 in week 10) and drastically lose focus afterwards (down to 0.11).
	The TV debate increased the tagging focus of the CDU/CSU, SPD, FDP, and the Greens, but did not gain enough attention by the Left and the Pirates to do so for them.
	The latter were the two parties which could not hope for being a junior partner in the coming government coalition. 

	The two events hardly caused changes of focus in the retweeting practice (cf. Figure \ref{RTEntr}). One can further see that the reweeting as well as the mentioning practice have a very low focus (high entropy) compared to the tagging practice and don't seem to be impacted by external events. 
	For the retweeting practice we observe a slight increase of focus during the TV debate. A closer look into our data reveals that the debate caused the SPD and Greens to retweet what their most vocal members had to say. The SPD's chairman S. Gabriel and the Green's parliamentary managing director V. Beck were even subject of their parties' strongest retweet bursts.
	These two parties' activities contributed most to slightly increasing the retweet focus on average.

	The absence of any change in the mentioning practice (cf. Figure \ref{@Entr}) reveals that, even if real world events like the debate or the election itself caused the average party to sharpen its thematic focus, it did so without changing its focus of who to talk to.
	
	\smallskip
	\textbf{Cultural Similarity:}
	Even though parties increase their individual thematic focuses towards the election, they don't become more similar to each other until the election debate causes them all to increasingly focus on overlapping hashtags (cf. Figure \ref{TagCos}).
	This general tagging behavior suggests that, absent punctuations, parties try to maintain distinctiveness by finding a topical niche and focusing on it.
	This time, the election debate also impacts the Left and Pirates and causes them to become more similar to other parties, though they remain least similar to other parties over time. 
	This suggests that all parties included topics which emerged during the TV debate into their tagging practice. 

	Interestingly, neither of the events had an impact on the similarities between parties according to their retweeting practices, but both impacted their mentioning similarities (cf. Figure \ref{RTCos} and \ref{@Cos}).

	One possible explanation for this observation is that the debate narrowed down the pool of politicians to those which were subject of the debate.
	Evidence for this explanation can be found by taking a closer look at the data: both parties whose candidates were debating (CDU/CSU and SPD) experienced their strongest bursts as their members started mentioning the SPD's candidate P. Steinbrück who also bursted strongly for the Greens and received most mentions that week from the FDP.
	The CDU/CSU candidate A. Merkel, who did not maintain a Twitter account, was tagged instead, her hashtag \#merkel bursting in week 7 for all parties (see Table \ref{facts} for details). This indicates that the TV debate caused changes in the mentioning and tagging practices of most parties, while neither the debate nor the election caused any party to notably start retweeting another party's voices.

	\smallskip
	\textbf{Cultural Reproduction:}
	So far, we have shown that parties increase their thematic focus towards the election and become more similar according to their tagging practices as the election comes closer and the debate bursts.
	However, it remains unclear how stably the parties reproduce their focuses.
	Recall that a style is absent if cultural reproduction decreases.

	Average tagging reproduction -- the stability of practice -- increases from 0.49 to 0.63 in election week 9 where it peaks and subsequently drops (cf. Figure \ref{TagStab}). The average party's stability increases, focuses are increasingly reproduced, and turnover is reduced.
	But the style is strongly perturbed by the TV debate.
	The sudden drop in reproduction means that the punctuation keeps parties, on average, from increasingly narrowing their focuses. They lose stability.
	Again, the Left and the Pirates are not affected by this event. But least effected is the SPD.
	This is because the SPD had the largest turnover of hashtags in the two weeks before the debate (i.e., the SPD was least stable in weeks 5 and 6, as one can see in Figure \ref{TagStab}). Consequently, the event did not punctuate their style.
	The CDU/CSU, on the other hand, had a strong reproduction of focus and was consequently negatively affected by the event.
	The debate does not have an impact on reproduction in retweeting (cf. Figure \ref{RTStab}) and mentioning (cf. Figure \ref{@Stab}).
	This is expected because focus and similarity were not, or only weakly, affected by the event.
	There is a slight trend in the retweet practice for all parties, except the FDP (cf. Figure \ref{RTStab}).
	Parties increasingly reproducing their retweeting focus, while not becoming more similar to each other (cf. Figure \ref{RTCos}), suggests that they maintain distinctiveness in terms of who their spokespersons are -- the homophily principle at work.

	But all growth of reproduction must come to an end and actually does after the election, as one can see in Figures \ref{TagStab}, \ref{RTStab}, and \ref{@Stab}. Election styles terminate and post-election styles start, from lower levels of reproduction.
	Average mentioning reproduction is noisy but rather constant over time, indicating that the debate about which politician is worth commenting on is progressing with smooth turnover.

	\smallskip
	\textbf{Institutionness:}
	Institutionness $I$ captures how long a fact is getting referenced often. Since election styles had a duration of up to 10 weeks, any fact with $I>10$ can be regarded to represent a defining factor for a party's identity.
	Indeed, among the facts with $I>10$, we find environment-related hashtags only for the Greens and internet and surveillance-related tags mostly for the Pirates.
	Unsurprisingly, each party's name hashtag (\#cdu, \#csu, \#spd, etc) has a score $I=13$.
	Table \ref{facts} shows that the election hashtag \#btw13 has the same score $I=13$ for all parties but the FDP ($I=10$).
	This points at the fundamental difference between the two events discussed above. The election is a long-term topic that endures and is repeatedly reproduced by long-term styles. The TV debate is a bursty punctuation interrupting these styles (see low institutionness and high burstiness scores for all parties in Table \ref{facts}).
	Politicians getting retweeted or mentioned for more than 10 weeks are all top-runners and authorities with functions like minister, vice chairman, or political managing director.

	\smallskip
	\textbf{Burstiness:}
	So far, there is no evidence that hashtags that were important online in the TV debate episode were actually new.
	But in fact, the average proportion of hashtags not referenced before week 7 is 65\% across parties, a percentage not reached in any other week.
	The increase of focus and similarity, and the decrease in reproduction in tagging are likely due to this renewal.

	Of all hashtags having its burst onset in week 7, due to the selection of the sample, only a handful used by the Pirates is devoid of political relevance or meaning.
	Table \ref{facts} shows examples of selected facts and their burstiness scores.
 	
	\section{Discussion}
	\label{discussion}

	In our empirical study, we found that the focus on hashtags is more concentrated and similarities between parties through common hashtag usage are much more pronounced than foci on, and similarities through, retweeted or mentioned politicians.
	Tagging is also more prone to perturbation by offline events than retweeting or mentioning.
	Retweet practices are much more stable because the users who enjoy attention tend to remain the same even as bursts of topical activity propagate through the system.
	Even though studying each practice is diagnostic of socio-cultural process, studying tagging is most indicative of culture as the use of language is the ultimate form on symbolic communication \cite[Ch. 4]{Padgett2012}.
	The notion of ``social'' is stronger in retweeting and mentioning, where the facts and the users referencing the facts are identical.

	Furthermore, we used our approach to study the effects of a major offline event, the TV debate of the two top candidates three weeks prior to the election. It left a distinct mark in the tagging practice of politicians but only impacted some aspects of their retweeting an mentioning practice.
	In addition to revealing differences between distinct practices we also found interesting differences between individual parties.
	For example, while most parties increased their tagging focused during the week of the TV election debate, the focus of the Pirates and the Left was not impacted by this event.
	This points at differences in what parties were trying to control: All but the latter were campaigning for becoming a part of the new government -- the latter never had a chance to be more than part of the opposition.
	The results of our empirical study can be seen as anecdotal evidence for the fact that our approach is suitable for detecting differences in the conversational practices of one or several groups of users.

	Regarding the objects of study, a typical dynamic for an average party was described: In the run-up to the election a party has a style of practice geared toward controlling uncertainties in communication. Certain topics are focused on and reproduced. Members of other parties are not retweeted but mentioned if an event -- in out case the TV election debate -- creates a need to do so.
	After the election, attention and control is directed in another direction. Style is altered but certain institutions -- well-established (self-)references in topics and users -- remain at the core of how parties define themselves.

	An interesting and plausible extension of this work would be the study of inter-party similarities rather than average similarities over longer periods of time.
	Finally, our computational approach in general can also be applied to understand how non-politicians participate in political conversations as well.

	\section{Conclusions}

	\label{conclusion}

	We have presented a computational approach to assessing online conversational practices of political parties on Twitter. Our approach is rooted in, and informed by, theoretical constructs from relational sociology, in particular cultural focus, - similarity, and - reproduction of agents as well as institutions and punctuations. We devised measures for each of these dimensions to enable the computational assessment of socio-cultural structure and dynamics of online conversational practices. We presented our approach and demonstrated its usefulness in a study on the German federal election 2013.

	Overall, we find that several online conversational practices differ significantly. We highlight these differences along with interesting commonalities among political parties by studying political communication on Twitter.

	While our computational approach was applied to a single case study, the approach is not limited to a single case. Rather it is general enough that it can be applied to other contexts in which different groups of agents 
	communicate and to other social media systems similar to Twitter.
	We hope that our work equips future computational social scientists with an improved instrument to assess and study online conversational practices in social media over time.

	\balance
	\footnotesize
	\bibliographystyle{aaai}
	\bibliography{arxiv}

\end{document}